\documentclass[reprint,amssymb,superscriptaddress,aps, prl, twocolumn]{revtex4-2} 
\usepackage{amsmath,amssymb,bm,ulem}
\usepackage{graphicx}
\usepackage[dvipsnames]{xcolor}
\usepackage[colorlinks=true,allcolors=blue, breaklinks]{hyperref}
\usepackage{xurl}
\usepackage{xr}
\usepackage{dcolumn}
\usepackage[capitalize]{cleveref}
\usepackage{bm}

\newcommand{\beq}{\begin{eqnarray}}
\newcommand{\eeq}{\end{eqnarray}}

\def \beq{\begin{eqnarray}}
\def \eeq{\end{eqnarray}}

\newcommand{\bq}{\bm{q}}

\begin{document}

\title{Nonreciprocal surface tension: anisotropy-induced defect motility and organization}
\author{Laya Parkavousi}
\email{laya.parkavousi@ds.mpg.de}
\author{Suropriya Saha}
\email{suropriya.saha@ds.mpg.de}
\affiliation{Max Planck Institute for Dynamics and Self-Organization (MPIDS), D-37077 G\"ottingen, Germany}

\date{\today}

\begin{abstract}
We show that interfacial nonreciprocity transforms defect dynamics in conserved scalar fields within the framework of the Nonreciprocal Cahn–Hilliard model. Nonreciprocal surface tension alone produces intermittently stable defects: system-spanning target patterns form, lose stability, self-destruct, and nucleate again from a defect-chaotic state. When bulk and interfacial contributions interplay in a particular way, the system forms a distinct mosaic-wave state: traveling waves remain coherent within finite domains demarcated by linear arrangements of motile dislocations, which act as lines of phase slip. Mosaic-waves exhibit scale-free fluctuations at length scales much larger than the average wavelength of the traveling patterns. To explain the wide range of emergent dynamics, we construct the dynamics of the Goldstone-mode. The nonlinearities governing its large-scale fluctuations belong to the anisotropic Kardar–Parisi–Zhang universality class, with the sign of the nonlinear anisotropy controlling the nature of the out-of-equilibrium dynamics.
\end{abstract}
\maketitle

Nonreciprocity, or action-reaction symmetry breaking, has emerged as a paradigm for breaking detailed balance at the microscopic level, on par with self-propulsion~\cite{ramaswamy2010mechanics,soto2014self,saha2019,shankar2022topological,golestanian2024non,fruchart2026nonreciprocal}. Its inclusion is associated with striking non-equilibrium dynamics in both theoretical~\cite{saha2020,pisegna2024emergent,DadhichiMaitra2020,Loos2023VisionConeXY,Dinelli2023,Du2025VisionConeXYLatticeGas,Bandini2025VisionConeXY,Ouazan-Reboul2023,YuBenoit2024,BraunsPRX} and experimental~\cite{chen2024emergent,guillet2025melting,tan2022odd,chao2026selective,SriramExptNRCH_PhysRevLett.133.208301,berezney2026active,Lavergne2019VisualPerceptionMotility} works. A common thread connecting nonreciprocal mixtures of two or more species, whether living or active, scalar or polar, is that these seemingly disconnected systems can break time translation symmetry and evolve into time-crystalline steady states. Oscillations can be system wide~\cite{fruchart2021} or associated with a finite length-scale~\cite{saha2020,you2020,ThieleTuring} depending on conservation laws, and may appear on crossing an exceptional point with a simultaneous breaking of parity and time-reversal (PT) symmetry. Often used interchangeably with non-hermiticity, the concept of nonreciprocity finds widespread applications across disciplines far beyond the realm of active matter, for example, in driven quantum systems~\cite{Ding2022NonHermitian}.  

Nonreciprocal scalar mixtures are minimally described by the Nonreciprocal Cahn--Hilliard (NRCH) model~\cite{saha2020,you2020}, several variants of which have been explored, including the role of nonlinearities~\cite{saha2025effervescence} and multiple species~\cite{parkavousi2025compositionaldisordermulticomponentnonreciprocal}, nonreciprocity in surface tension has remained largely unexplored despite naturally emerging in coarse-grained theories~\cite{tucci2024nonreciprocal, sahoo2025nonreciprocalmodelbrole,YuBenoit2024}. NRCH has been established as a framework to explore pattern formation in number-conserving systems~\cite{ThieleNonreciprocal_Classes} and is useful for describing systems such as mass-conserving reaction-diffusion systems~\cite{Frey-prx-2020,BraunsPRX}, quorum-sensing particles~\cite{YuBenoit2024}, and chemically active systems~\cite{Ouazan-Reboul2023}. The central theme of this work is defect dynamics in the NRCH model with the inclusion of non-reciprocity in the surface tension. Defect dynamics assumes fascinating forms in active matter, from variants of the Kuramoto model to defect cloaking in spins with vision cone~\cite{Popli2025DefectCloaking,Rouzaire2022KuramotoDefects,Rouzaire2025NonreciprocalDefects,myin2026breakdownemergentchiralorder,woo2026extensivespatiotemporalchaosnonreciprocal}. We find that interfacial nonreciprocity, both alone and through its interplay with bulk nonreciprocity, gives rise to distinct defect dynamics linked to anisotropic KPZ dynamics~\cite{Wolf1991} and defect unbinding~\cite{Sieberer2016,Wachtel2016}.

\begin{figure*}[t]
    \centering
    \includegraphics[width=0.95\linewidth]{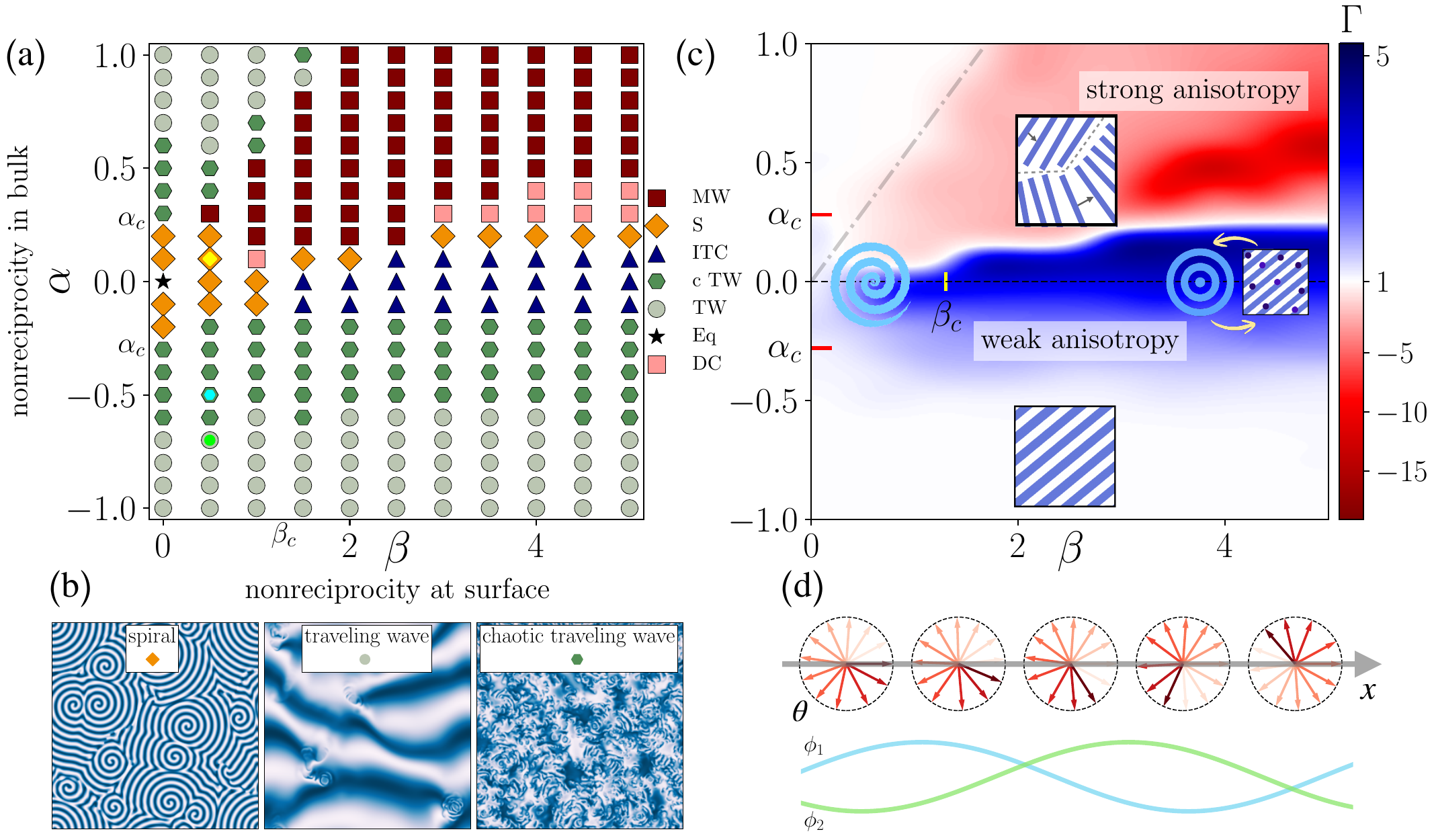}
    \caption{ \textbf{Nonreciprocal surface tension: defect dynamics and anisotropic KPZ.} Emergent dynamics in the Nonreciprocal Cahn-Hilliard model is explored by varying the strength of activity in the bulk ($\alpha$) and at the interface ($\beta$). (a) A variety of steady states with distinct defect dynamics is observed. Two hitherto unreported states, intermittent target chaos (ITC) and Mosaic-waves (MW), arise from the interplay of $\alpha$ and $\beta$ [see Figs.~\ref{fig:fig2} and \ref{fig:fig3}]. Panel (b) depicts some dynamical states that are also found for $\beta=0$. (c) The asymmetry between the top and bottom halves of (a) is explained by the phase dynamics. The anisotropy coefficient $\Gamma$, defined as the ratio of the KPZ nonlinear couplings parallel and perpendicular to the direction of wave propagation, varies across parameter space. A change in its sign marks the transition from isotropic spirals to MW, and from MW to traveling waves with a discontinuous jump in wavenumber (gray dashed line). (d) Fluctuations around oscillatory states are described by phase dynamics governed by KPZ nonlinearities. For nonzero $\beta$, these nonlinearities become anisotropic, producing qualitative changes in defect dynamics.}
    \label{fig:fig1}
\end{figure*}

\begin{figure*}[t]
    \centering
    \includegraphics[width=0.9\linewidth]{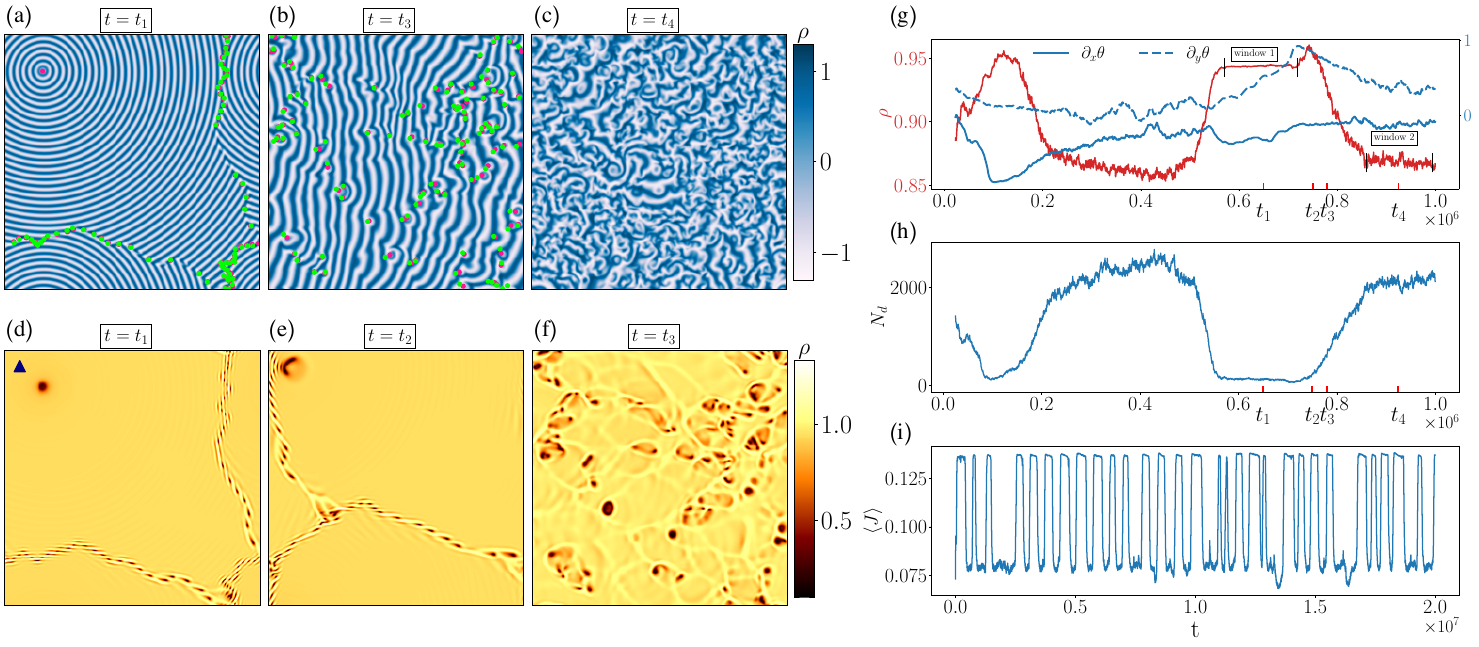}
    \caption{ \textbf{Intermittent target chaos}: Heatmaps of the number density of species 1 and of the amplitude $\rho$, a function of both fields, are shown in panels (a--c) and (d--f), respectively, at times $t_{1-4}$ within a cycle, for $\alpha=0$ and $\beta=4$. Panel (a) shows a single system-spanning target, with waves emanating from a point. The green and red dots mark positive and negative defects; the stable target carries charge $+1$. At the defect core, shown in panel (d), $\rho$ is close to zero. Panel (e) shows the distorted core and motile boundary shortly before collision. After the collision, the target is replaced by traveling waves and several bound defect cores of opposite signs, panel (b). The traveling wave is then destroyed by target proliferation, one of which evolves into another transiently stable target, and the cycle repeats, panel (c). (g) Spatially averaged $\rho$, $\partial_x\theta$, and $\partial_y\theta$ reflect this evolution, panel. The plateau in $\rho$ corresponds to the stable target, while the peak occurs when traveling waves with defects are formed and decays to a lower value in the spatio-temporally chaotic state. (h) The density of target cores follows a similar trend, remaining small when the target is stable and increasing during chaotic dynamics. (i) The process continues indefinitely as a statistical steady state, with the average value of $\bm{J}$ cycling repeatedly.   }
    \label{fig:fig2}
\end{figure*}

\begin{figure}[t]
    \centering
    \includegraphics[width=1\linewidth]{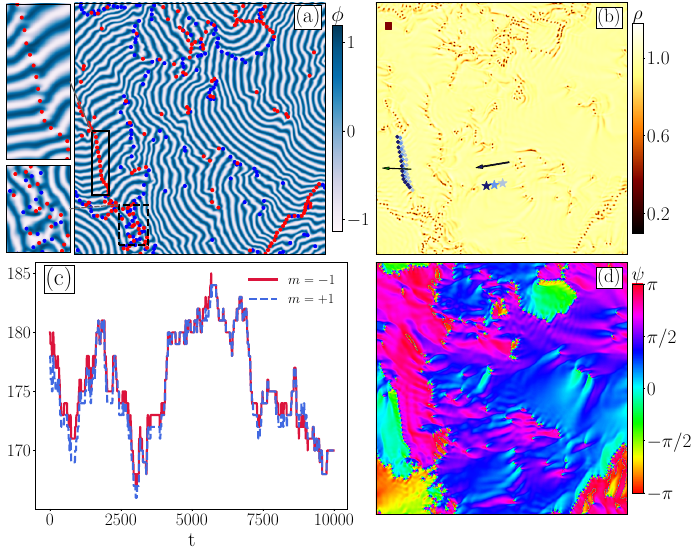}
    \caption{ \textbf{Mosaic-waves with long-lived dislocations and slip-lines}: (a) The heatmap shows a state of dislocation chaos, for $\alpha=0.6$ and $\beta=4$. The solid black box highlights a line of defects, while the dashed black box indicates a cluster of defects that remain close to one another. (b) Heatmap of the amplitude of the field, $\rho$, which approaches zero at the defect cores. On the left side of the box, drifting lines of defects are visible; the arrow indicates the direction of their evolution. (c) Dislocations are topological defects that are tracked in the simulation, and both their creation and annihilation occur pairwise. Total number of $+1$ and $-1$ dislocations as a function of time. (d) The local direction of propagation along $\psi = \cos^{-1} (\hat{\bm{x}} \cdot \bm J) |\bm J|^{-1} $ is plotted as a heatmap shows discontinuous changes in $\psi$.}
    \label{fig:fig3}
\end{figure}

{\it{Summary.---}} The inclusion of two nonreciprocity coefficients, $\alpha$ and $\beta$ (representing bulk and interfacial contributions, respectively), produces a variety of dynamical states (see Fig.~\ref{fig:fig1}(a)), including new forms of self-organization that we term intermittent target chaos (ITC) and Mosaic-waves (MW). These are observed in addition to stable spirals and chaotic traveling waves reported previously in NRCH, see~\cite{rana2024defectPRL,rana2024defect}, and Fig.~\ref{fig:fig1}(b). In ITC, targets are transiently stable before self-annihilating, as the system evolves into a state of defect chaos with several target cores nucleating and disappearing, see Fig.~\ref{fig:fig2}. Subsequently, a new stable target nucleates, and the roughly bistable process (steady targets and defect chaos) repeats indefinitely. For nonzero $\alpha$ and $\beta$, the relative signs of the coefficients determine the final state of the system - for $\alpha,\beta>0$ we find mosaic-waves (MW), where the defects self-organize to form domains with directed traveling waves as shown in Fig.~\ref{fig:fig3}. The form of the state diagram, in particular the asymmetry between the top and bottom halves in Fig.~\ref{fig:fig1}(a), piques interest, a result that we explain by examining the phase dynamics.

All defect-ridden states, including ITC and MW, originate in a region where traveling waves are linearly stable, highlighting their nonlinear origin. Oscillatory NRCH patterns resemble spatially coupled synchronized nonlinear oscillators, with a compact phase on $(0,2\pi)$ set by the relative densities of the two species [Fig.~\ref{fig:fig1}(d)]. For traveling density waves, represented in space by blue and green lines for species $1$ and $2$ in Fig.~\ref{fig:fig1}(d), the phase varies linearly in space and time, and the amplitude, radius of the circle in Fig.~\ref{fig:fig1}(d), plays the role of the order parameter for the PT-symmetry-breaking transition~\cite{saha2020,pisegna2024emergent}. The descriptions holds when amplitude fluctuations relax on finite timescales and can be enslaved, such that the Nambu-Goldstone mode represented by a unit vector evolving on the simplest limit cycle of two spatially coupled fields~\cite{pisegna2024emergent,daviet2025kardar}. Its physical meaning is system-dependent, describing, for example, smectic undulations~\cite{chen2024emergent,ChenTonerPolarSmectics}, the macroscopic phase of an exciton-polariton condensate~\cite{Vercesi2023}, or orientation in chiral liquid crystals~\cite{maitra2020chiral,maitra2025activity}. However, its dynamics follow from symmetry and generic non-equilibrium conditions, with universal behavior captured by the Kardar--Parisi--Zhang (KPZ) equation, originally formulated for scale-free interface fluctuations~\cite{KPZ1986} and now recognized as broadly relevant~\cite{widmann2026observation}. We show that $\beta$ and $\alpha$ together render these phase fluctuations dependent on the polarity of the traveling waves, such that the emergent states are governed by large-scale anisotropic phase fluctuations around stable traveling waves, which qualitatively alter defect dynamics~\cite{Altmann_PhysRevLett.121.085704, Altmann_PhysRevX.5.011017}.

{\it{Theoretical model.---}} The conserved number densities of species $1$ and $2$, $\phi_{1,2}$, are combined into a complex field $\phi = \phi_1 + i \phi_2$ which evolves as
\beq 
\partial_t \phi = \nabla^2 \left[ -(1+i\alpha) \phi + |\phi|^2 \phi - (1+i \beta) \nabla^2 \phi \right].
\label{eq:PhiDyn}
\eeq 
The equations represent a conserved version of the complex Landau-Ginzburg dynamics, with the important difference that the term $\alpha$ represents a global rotation and is thus irrelevant for the non-conserving dynamics. Active terms proportional to coefficients $\alpha$ and $\beta$ represent non-variational contributions, while the passive part is derivable from a total free energy $F = \int_{V} \mbox{d}^2\bm{r} f(\bm r)$ which a volume integral of the free energy density $f(\bm{r})=  -|\phi|^2/2 + |\phi|^4/4  + |\bm{\nabla} \phi |^2$, a standard choice to enable analytical approaches later in the paper~\cite{saha2025effervescence,pisegna2024emergent}. The system reduces to passive model B describing bulk-phase separation for $\alpha=\beta=0$ showing a transition to traveling states for nonzero $\alpha$ or $\beta$. A systematic coarse-graining of mixed suspensions of Janus-colloids or quorum-sensing particles yields the NRCH model, where asymmetry manifests at all orders in the spatial-gradient expansion~\cite{tucci2024nonreciprocal,YuBenoit2024}. The second- and fourth-order nonreciprocal terms represent deviations from free-energetic bulk and interfacial contributions to the chemical potential, respectively. 

We define a few quantities used to characterise the dynamical states: the amplitude $\rho = \sqrt{\phi_1^2+\phi_2^2}$ and the phase $\theta = \tan^{-1} (\phi_2 \phi_1^{-1})$ such that $\phi = \rho \exp(i \theta)$. Finally, we define a quantity related to both $\rho$ and $\theta$, an asymmetric combination of the gradients of the number densities, given by $\bm{J} = (\phi_1 \bm{\nabla} \phi_2 -  \phi_2 \bm{\nabla} \phi_1) \equiv \rho^2 \bm{\nabla} \theta $. $\phi = \rho_0 \exp (i \bm{q}_0 \cdot \bm{r} - \Omega t)$, where $\rho_0^2 = 1 - q_0^2$, and $\Omega = (\alpha-\beta q_0^2)q_0^2$ are a class of exact solutions of Eq.~\eqref{eq:PhiDyn}, parametrized by the wavenumber $q_0$ that can assume values in the closed interval $(0,1)$. It suffices to consider positive $\beta$ only while $\alpha$ can have either sign because if the signs of both parameters are simultaneously reversed, $\phi^*$ (instead of $\phi$) is a solution of Eq.~\eqref{eq:PhiDyn}. 
\\\\
{\it{State diagram.---}} The steady-state collective behavior is mapped out by carrying out simulations with $\alpha$ varied over the range $(-1,1)$ in intervals of $0.1$ and $\beta$ over the range $(0,5)$ in intervals of $0.5$, see Fig.~\ref{fig:fig1}(a). See Appendix~\ref{app:secB} for the measures used to quantify the states. For $\alpha = 0$, and $\beta$ lower than a threshold $\beta_c$, multiple coexisting spirals are observed. On increasing $\beta$ beyond $\beta_c$, the system undergoes cycles of stable targets and defect chaos in the steady state called Intermittent target chaos (ITC). The transition roughly mirrors a transition from a defect ridden state to steady traveling waves for  which $\beta = 0$ and $\alpha$ nonzero, where spirals are unstable beyond a threshold value $\alpha_c$, see Fig.~\ref{fig:fig1}(a). The traveling waves beyond this point are chaotic for values close to $\alpha_c$, becoming more and more stable with increasing $\alpha$. For $\alpha$ and $\beta$ of opposite signs, the steady-state dynamics is largely controlled by $\alpha$; irrespective of the value of $\beta$, similar dynamical trends are observed. When the signs of $\alpha$ and $\beta$ are the same, the steady state is strikingly different - the system evolves to a state formed of patches of traveling waves moving in different directions which we call mosaic-waves (MW).

{\it{Intermittent target chaos (ITC).---}} The system is intermittently chaotic for $|\beta| \gg |\alpha|$, undergoing dynamical processes that repeat in an approximately periodic manner, see Movie S1~\cite{SI}. A single system-spanning target emerges and persists for a finite time: circular traveling waves emanate from a circularly symmetric center where both fields vanish, such that $\rho = 0$ at this point, as shown in Figs.~\ref{fig:fig2}(a,d). The target carries a topological charge of $+1$, as calculated by the line integral of $\bm{J}$, $\oint \bm{J}\cdot \mbox{d}\bm{r}$ around a path encircling the point. This charge is balanced by an irregular boundary composed of anisotropic, and hence motile, defects.

The boundary advances over timescales slow compared to the period of the emitted waves, eventually reaching the defect core. Interaction between the core and the boundary distorts the initially circular core, which loses stability and gives way to traveling waves that move roughly in the same direction before being destroyed by randomly generated defect cores, leading to a nearly isotropic defect-chaotic state. The cycle repeats when one defect core coarsens at the expense of the others and evolves back into a single target. These processes are reflected in the dynamics of the average $\rho$, and the directional derivatives $\partial_x \theta$ and $\partial_y \theta$ of the phase, shown in Fig.~\ref{fig:fig2}(g). $\rho \approx 0.95$ when the target exists, falling to values close to $0.85$ when the system is filled with defects. $\partial_{x,y}\theta$ increases when the waves propagate predominantly along the $x$ or $y$ direction, quantifying the anisotropy. Fig.~\ref{fig:fig2} also shows the rise and fall in the number of topological defects during the same time window. Finally, Fig.~\ref{fig:fig3} shows the periodic rise and fall in the magnitude of $\bm J$, averaged over the system size and plotted over a long enough time interval such that the system undergoes $20$ or more cycles.

{\it{Mosaic-waves (MW).---}} For a given $\alpha$, and a sufficiently large value of $\beta$ with the same sign, the system evolves into a mosaic structure consisting of domains within which the layered structure is preserved, see Fig.~\ref{fig:fig3} and Movie S2~\cite{SI}. Within each domain, the layers translate along the layering direction. The domains are stabilized by arrays of dislocations of the same sign.

The defects always form and annihilate in pairs, thus conserving the total topological charge as they should. Defects carrying the same charge segregate to form the domain walls, which allow the traveling waves to slip past one another. In contrast to ITC state, there are large domains that are free of defects. The defects persist in the steady state; their total number fluctuates with time around a steady constant value, see Fig.~\ref{fig:fig3}(c). Defects distort the density layers as seen in Fig. \ref{fig:fig3}(a). The heatmap of $\rho$ shows ripples corresponding to large scale deformations of the density layers, see Fig.~\ref{fig:fig3}(b) and Movie S2~\cite{SI}. The static structure factor defined as the temporally averaged density $S(q_x,q_y) = |\phi(q_x,q_y)|^2$ shows anisotropy, as seen in Fig.~\ref{fig:fig3}(d). 

{\it{Phase dynamics.---}} To explain the asymmetry in  dynamics for same and opposite signs of $\alpha$ and $\beta$, see Fig.~\ref{fig:fig1}(a), we now turn our attention to phase fluctuations $\delta \theta$ around the traveling wave with $\bm{q}_0 = q_0 \hat{\bm{x}}$ and construct its dynamics. An adiabatic elimination of perturbations in the amplitude, $\rho$, (see Appendix~\ref{app:secA}) leads to the following equation for $\delta \theta$, 
\beq
&& \partial_t\delta\theta
=V\,\partial_x\delta\theta
+D_L\,\partial_x^2\delta\theta
+D_T\,\partial_y^2\delta\theta \nonumber \\
&& +\lambda_1(\partial_x\delta\theta)^2
+\lambda_2(\partial_y\delta\theta)^2,
\label{eq:akpz_app}
\eeq
$\delta \theta$ is advected by the velocity $V$, and it diffuses in the longitudinal and transverse direction with diffusion coefficients $D_L$ and $D_T$, respectively, see Appendix~\ref{app:secA} for their full expressions. The lowest order and relevant nonlinearities shown in the second line in Eq.~\eqref{eq:akpz_app} are those of the aforementioned compact Kardar-Parisi-Zhang~\cite{KPZ1986} equation, and are proportional to coefficients $\lambda_1=-\alpha+6\beta q_0^2$ and $\lambda_2=-\alpha+2\beta q_0^2$ in $x$ and $y$ directions. As explained before, it defines a universality class for systems where diffusive smoothing (given by $D_{L,T}$) competes with nonlinear growth that enhances fluctuations. The KPZ terms in Eq.~\eqref{eq:akpz_app} break $x \to -x$, and $\delta \theta \to -\delta \theta$ symmetries simultaneously, thereby revealing the active nature of the fluctuations. Consequently, these terms are proportional to the two nonreciprocal parameters. Transforming to a co-moving frame, and scaling the $x$ and $y$ directions using the diffusivities $D_L$ and $D_T$ we can write the equation for $\delta \theta$ as $\partial_t \delta \theta = \nabla^2 \delta \theta + \lambda_L (\partial_x \delta \theta)^2 + \lambda_T (\partial_y \delta \theta)^2$ thus defining the anisotropy coefficient $\Gamma$ 
\beq 
\Gamma = \frac{(\alpha-6\beta q_0^2)(1-q_0^2)}
{(\alpha - 2\beta q_0^2)(1-3q_0^2)}.
\label{eq:anisotropy}
\eeq
$\Gamma=1$ represents an isotropic system, implications of which were discussed in~\cite{pisegna2024emergent}, while deviations from unity show that phase fluctuations at large length-scales are anisotropic.

The degree of anisotropy is determined jointly by $\beta$, and $q_0^2$, the latter of which is a dynamical quantity, as opposed to the former, which is a fixed parameter. $q_0$ exhibits multi-stability, so a wide range of values is accessible, and the value realized in numerical solutions of Eq.~\eqref{eq:PhiDyn} is selected by the initial conditions and nonlinearities~\cite{CrossTu1994}. Conservation of particle densities ensures that $q_0$ remains finite, creating the possibility that $\Gamma$ can change sign when $\alpha$ and $\beta$ are both positive. For a constant $q_0$, this happens in two ways, when $\lambda_1$ and $\lambda_2$ changes sign when $\alpha/\beta$ equals $6q_0^{2} $ and $2q_0^{2} $ respectively, meaning that $\Gamma$ is negative in the region bound by the two lines with different slopes. $\Gamma$ is shown in the heatmap spanning $\alpha$–$\beta$ in Fig.~\ref{fig:fig1}(c), it is indeed negative in large parts of the upper half plane and well approximated by a constant $q_0$. 

Perturbative renormalization group analysis of anisotropic KPZ predicts that in two dimensions, the interface enters a roughening regime with scale-free fluctuations at small wavenumbers for $\Gamma>0$, whereas for $\Gamma<0$ the nonlinearities are irrelevant~\cite{Wolf1991,tauber2002universality}. More importantly for the present results, the two regimes exhibit markedly different defect dynamics~\cite{Altmann_PhysRevLett.121.085704,Altmann_PhysRevX.5.011017,Wachtel2016,Szymanska_Vorten_anKPZ}. For $\Gamma>0$ (weak anisotropy), the nonlinearities screen defect interactions, allowing them to unbind and proliferate at all noise strengths, whereas for $\Gamma<0$ (strong anisotropy), the system flows toward equilibrium-like XY behavior, with long-ranged Berezinskii--Kosterlitz--Thouless-like defect interactions and defect proliferation only when entropy overcomes interaction energy~\cite{Berezinskii1971,Kosterlitz1974,KosterlitzThouless1973,Kosterlitz2016}.

We discover a strong correlation between the nature of the phase fluctuations and the defect dynamics by comparing panels (a) and (b) of Fig.~\ref{fig:fig1}. Regions where the phase fluctuations are strongly anisotropic are associated with Mosaic-waves, whereas weak anisotropy correlates with defect dynamics dominated by spirals or targets, which are either indefinitely stable or undergo intermittent cycles of destruction and re-nucleation. At the transition from blue to red, spirals lose stability to Mosaic-waves. On crossing the region from red to white, marked by a line in Fig.~\ref{fig:fig1}(c), a transition to traveling waves with a smaller $q_0$ is observed. For $\alpha = 0$ and $\beta \neq 0$, $\Gamma \sim 3$, placing the dynamics along this line within the weakly anisotropic regime, and providing some justification for the difference between the dynamics along the $\alpha \neq 0$, $\beta = 0$ and $\alpha = 0$, $\beta \neq 0$ axes.

Despite their differences, both the ITC and MW phases exhibit features consistent with anisotropic phase fluctuations leading to motile defects, in contrast to those reported previously~\cite{rana2024defectPRL}. For weak anisotropy, targets lose stability due to their moving boundaries, while spirals remain stable. Defects in the MW phase do not annihilate because they remain in motion~\cite{Chatterjee2021Inertia}; presumably, they could annihilate if they moved in random directions, as they do at the beginning of the simulations.

{\it{Discussion---}} To conclude, we find that a broad range of defect dynamics originates from nonreciprocal surface tension, revealing a hitherto unexplored facet of nonreciprocal mixtures. The phase dynamics in Eq.~\ref{eq:akpz_app} is the most general form allowed by symmetry, implying that higher-order terms in the number-density dynamics primarily renormalize the coefficients without qualitatively altering its structure. Our observations are consistent with the proposal that anisotropies can stabilize driven condensates that would otherwise be destabilized by defect proliferation~\cite{Altmann_PhysRevX.5.011017}. Yet, despite the central importance of KPZ dynamics, the consequences of its anisotropic form remain comparatively less understood. Our study therefore provides a starting point for further theoretical analysis of defect dynamics in anisotropic nonequilibrium systems. Since defects are also widely observed in mass-conserving reaction-diffusion systems, with quorum-sensing mixtures likely providing further examples, our framework may offer insight into the interactions between such defects~\cite{Denk2018MinE,Wettmann2018Geometry,Tan2020TopologicalTurbulence}.

\begin{acknowledgements}
We acknowledge insightful discussions with Ramin Golestanian and Navdeep Rana. We thank Ananyo Maitra for a thorough reading of the manuscript. We acknowledge support and funding from the Department of Living Matter Physics in MPIDS.
\end{acknowledgements}

\bibliography{biblio1}

\newpage
\onecolumngrid{}

\vspace{12pt}
\noindent\hrulefill \hspace{24pt} {\bf End Matter} \hspace{24pt} \hrulefill
\vspace{12pt}


\twocolumngrid{}



\renewcommand{\thesection}{\Alph{section}}
\setcounter{section}{0}

\refstepcounter{section}
\section*{Appendix \thesection}
\label{app:secA}
\noindent
We consider the NRCH equation for the complex field $\phi(\bm{r},t)=\phi_1+i\phi_2$ in Eq.~\eqref{eq:PhiDyn}, incorporating diffusive, nonlinear, and higher-order gradient terms with non-reciprocal parameters $\alpha$ and $\beta$. A plane-wave solution of the form $\phi=\rho e^{i\theta}$ is used as the base state, with phase $\theta=q_0 x-\Omega t$, amplitude $\rho_0^2=1-q_0^2$, and dispersion relation $\Omega=\alpha q_0^2-\beta q_0^4$. Small perturbations are introduced via $\rho=\rho_0+\epsilon\,\delta\rho$ and $\theta=q_0 x-\Omega t+\epsilon\,\delta\theta$, and the dynamics are expanded order by order in $\epsilon$.

\begin{figure*}[t]
    \centering
    \includegraphics[width=0.9\linewidth]{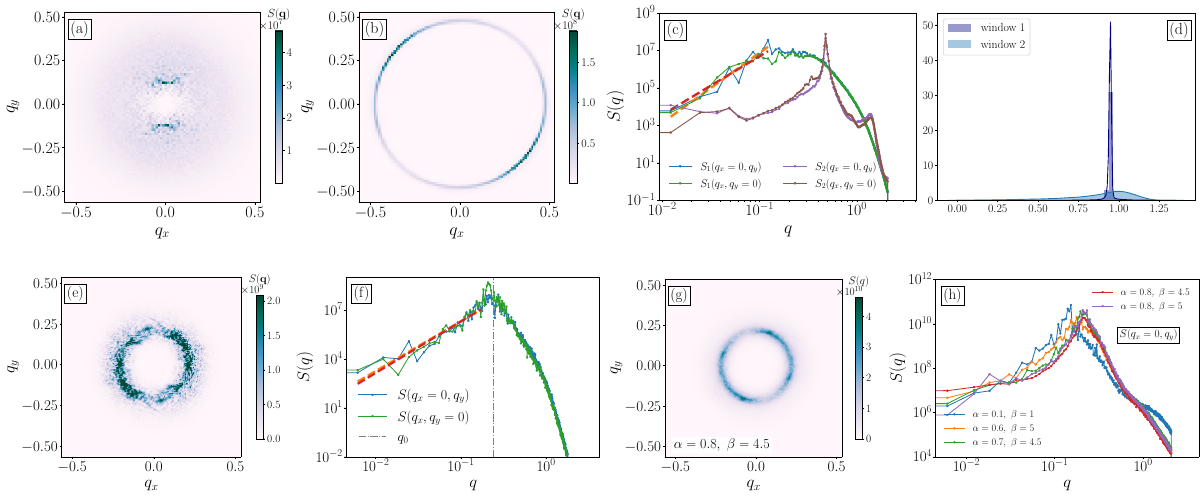}
    \caption{\textbf{Quantifying fluctuations in ITC and MW}: (a) The static structure factor $S(q_x,q_y)$ in the second window in Fig.~\ref{fig:fig2}(g). (b) The same as in (a) in the first window in Fig.~\ref{fig:fig2}(g). (c) S along the x and y axes. (d) The probability distribution of the amplitude is sharply peaked in window I and broad in II, encoding the fluctuations. (e) $S(q_x,q_y)$ for MW, and plots along x and y axes in the wavenumber plane in (f).}
    \label{fig3_appendix}
\end{figure*}

At linear order, the amplitude fluctuation $\delta\rho$ is stable and rapidly relaxes. It becomes slaved to gradients of the phase field, yielding the quasi-static relation
\begin{equation}
\delta\rho \approx -\frac{q_0}{\rho_0}\partial_x \delta\theta
+ c\,\nabla^2 \delta\theta.
\end{equation}
This reduces the problem to an effective phase-only description. Substituting this relation into the phase dynamics produces a linear anisotropic equation:
\begin{equation}
\partial_t \delta\theta =
V\,\partial_x \delta\theta
+ D_L\,\partial_x^2 \delta\theta
+ D_T\,\partial_y^2 \delta\theta,
\end{equation}
where $V$ is a drift velocity, and $D_L$, $D_T$ are anisotropic diffusion coefficients as
\begin{equation}
V=-2q_0(\alpha-2\beta q_0^2),
\qquad
D_T=q_0^2,
\qquad
D_L=\frac{q_0^2(1-3q_0^2)}{1-q_0^2}.
\label{eq:VD_app}
\end{equation}
Notice that $D_{L,T}$ are both positive in the range $(0,1/\sqrt{3})$. See Fig.~\ref{fig3_appendix}(d) showing the distribution of $\rho$ which peaks at values smaller than the threshold of the Eckhaus instability, implying that the phases in Fig.~\ref{fig:fig1}(a) arise from nonlinearities rather than a linear instability, in contrast to Ref.~\cite{saha2025effervescence}. At second order in fields, nonlinear gradient terms emerge, leading to an anisotropic Kardar--Parisi--Zhang (KPZ) equation:
\begin{equation}
\partial_t \delta\theta =
V\,\partial_x \delta\theta
+ D_L\,\partial_x^2 \delta\theta
+ D_T\,\partial_y^2 \delta\theta
+ \lambda_1(\partial_x \delta\theta)^2
+ \lambda_2(\partial_y \delta\theta)^2,
\end{equation}
with nonlinear coefficients
\begin{equation}
\lambda_1=-\alpha+6\beta q_0^2, \qquad
\lambda_2=-\alpha+2\beta q_0^2.
\end{equation}
Transforming to a co-moving frame and rescaling space and time yields the canonical anisotropic KPZ form:
\begin{equation}
\partial_\tau \delta\theta =
\nabla^2 \delta\theta
+ \lambda_L(\partial_X \delta\theta)^2
+ \lambda_T(\partial_Y \delta\theta)^2.
\end{equation}
A key dimensionless measure of anisotropy is
\begin{equation}
\Gamma\equiv \frac{\lambda_T D_L}{\lambda_L D_T}
=\frac{(\alpha-6\beta q_0^2)(1-q_0^2)}
{(\alpha-2\beta q_0^2)(1-3q_0^2)}.
\end{equation}
In summary, the NRCH model reduces to an anisotropic KPZ equation for the phase field. The non-reciprocal parameters $\alpha$ and $\beta$ control drift, anisotropic diffusion, and nonlinear growth, establishing a direct connection between microscopic nonreciprocity and macroscopic surface dynamics.


\refstepcounter{section}
\section*{Appendix \thesection }
\label{app:secB}
\noindent

\onecolumngrid
\twocolumngrid

In order to obtain the phase diagram, we used two different measures. The first measure is the spatial average of the order parameter $J$ for different values of $\alpha$ and $\beta$. This measure gives a good criteria to find the group of MW sharply, and also finding spirals and ITC in our phase diagram. The second measure is the parameter length that we calculate as

\begin{equation}
    q_0 = \frac{\sum S(q_x,q_y)\sqrt{q_x^2+q_y^2}}{\sum S(q_x,q_y)}
\end{equation}

The measure $q_0$ can detect the traveling waves in our phase diagram very effectively.

\begin{figure}[t]
    \centering
    \includegraphics[width=0.99\linewidth]{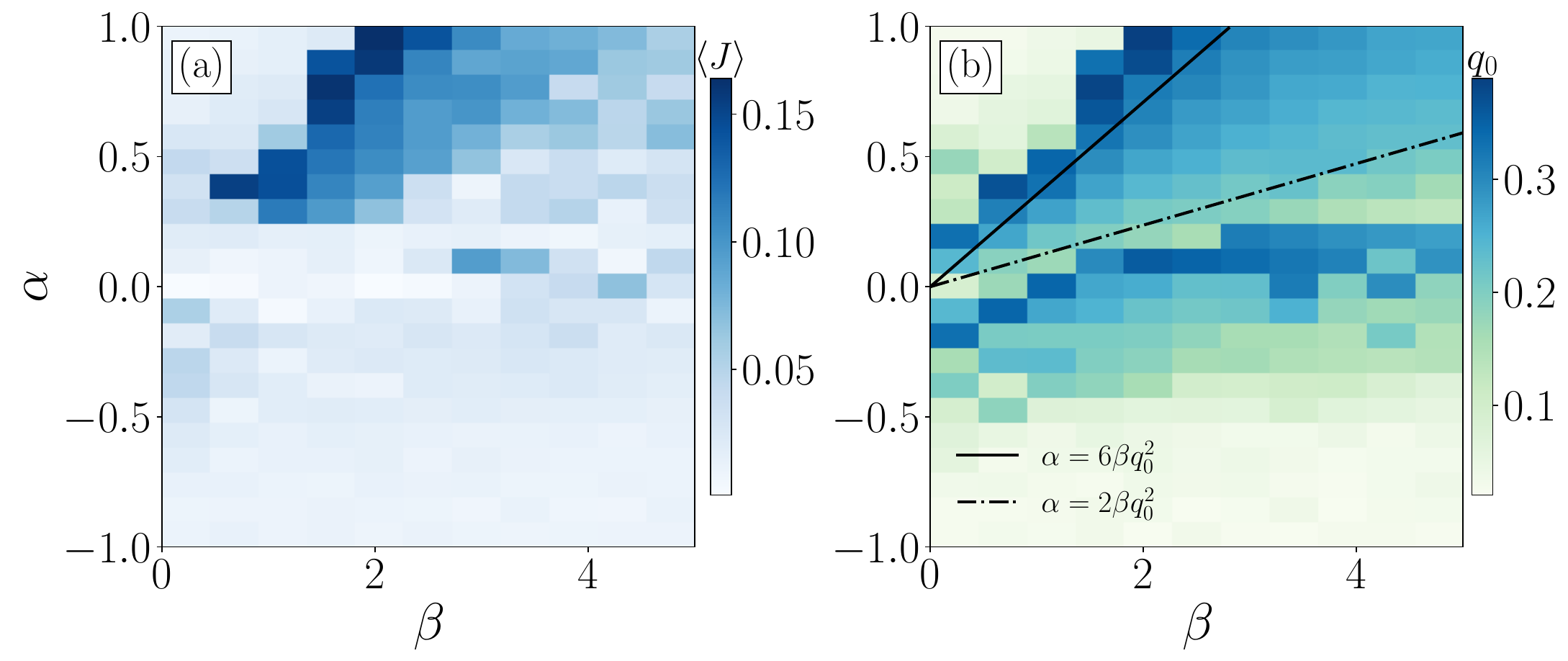}
   \caption{ \textbf{Order parameter and wavenumber selection}: (a) The average value of the spatially averaged order parameter $\bm J$, denoted by $\langle J \rangle$, changes discontinuously from large to small values, marking the transition from MW to traveling waves; see Fig.~\ref{fig:fig1}(a) for a comparison. (b) Heatmap showing the selected $q_0$ in the $\alpha$--$\beta$ plane, which, through the expression for $\Gamma$ in Eq.~\ref{eq:anisotropy}, results in the heatmap shown in Fig.~\ref{fig:fig1}(c). }
    \label{fig1_appendix}
\end{figure}


\begin{figure}[t]
    \centering
    \includegraphics[width=0.99\linewidth]{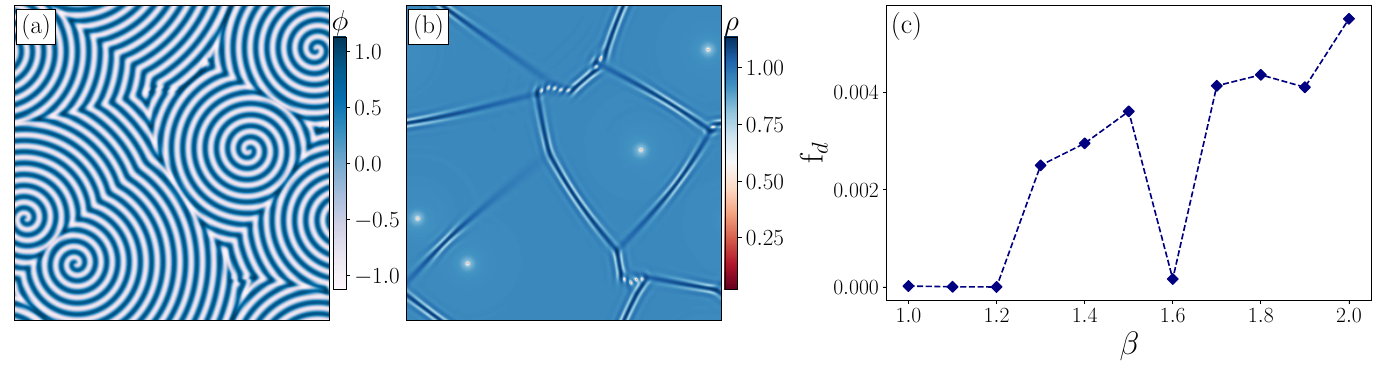}
    \caption{ \textbf{Transition from spirals to ITC}: (a) A stable spiral configuration (b) The $\rho$ field shows that the boundaries between the defects are well defined and stationary in sharp contrast to that in Fig.~\ref{fig:fig2}(d-f). (c) The average $J$ vanishes when the defects are stable, changing continuously to large values. It serves as an order parameter to determine the threshold $\beta_c$. }
    \label{fig2_appendix}
\end{figure}

\clearpage

\onecolumngrid 
\normalfont
\begin{center}
\textbf{\Large Supplemental Material: Nonreciprocal surface tension: anisotropy induced defect motility and organization}
\end{center}
\vspace{2ex}

The Supplemental Material contains details of the stability analysis of the equations of motion for vanishing average number densities of the two species, and of the numerical procedure followed to solve them including the nonlinearities. We also include further details of the states and information on the Supplemental Movies.

\maketitle
\section{Linear stability of the uniform states}
\noindent
To analyse the linear stability of steady homogeneous states $\boldsymbol{\phi}(x)=\left(\phi_1(x), \phi_2(x)\right)=\left(\phi_0, \phi_0\right) \equiv$ $\phi_0$ we perturb the densities slightly away from the uniform states as
\beq
\phi(\bq, t)=\phi_0+\varepsilon \phi() e^{i q x+\lambda t}
\eeq
Inserting the above into Eqs. 1 of the main text and linearizing in $\varepsilon \ll 1$ yields
\begin{equation}
\begin{aligned}
& \partial t \binom{\delta \phi_1}{\delta \phi_2}=-q^2 \mathbb{M} \cdot \binom{\delta \phi_1}{\delta \phi_2} =-q^2 \Tilde{M} \overline{\delta \phi}.
\end{aligned}
\end{equation}
 where the matrix $M$ and matrix $K$ are as follows,
\begin{equation}
M=\left(\begin{array}{cc}
1 & \alpha + \beta q^2 \\
-\alpha - \beta q^2 & 1
\end{array}\right)
\end{equation}
We want to study the trace and the determinant of the matrix $\Tilde{M} = -1-q^2$ and discriminant $\delta=-4\left(\alpha+\beta q^2\right)^2$
The eigenvalues $\lambda_{\pm}$ are given by
\beq 
\lambda_{\pm} = -\frac{q^2(1-q^2)}{2} \pm 2i q^2 |(\alpha + \beta q^2)|
\eeq 
Thus $\beta$ plays a role similar $\alpha$, with the difference that the complex part of $\lambda_{\pm}$ grow as $q^4$ (not $q^2$). Notice that for all parameters other than nonzero $\beta$ and $\kappa$, the complex parts grow as $\sqrt{|\beta^2-\kappa^2|}$, signalling an exceptional point at $\beta = \kappa$ occurring at a higher power of $q$.  
\begin{figure}[h!]
    \centering
    \includegraphics[width=0.8\linewidth]{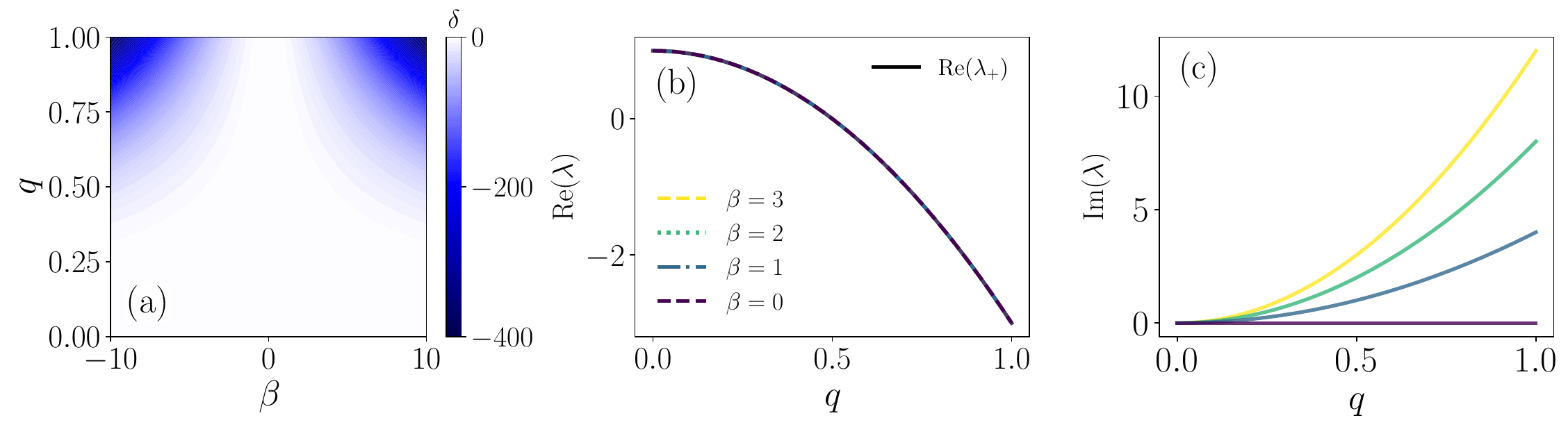}
    \caption{Figure showing the value of $\delta$ as a function of $\beta$ and $q$ illustrating the nature of the linear instability of the uniform composition that leads to pattern formation. (a) $\delta<0$ for $\beta \neq 0$ (all other parameters set to zero), meaning that complex eigenvalues arise at the smallest possible values of $q$ accessible. (b) A non-zero $\chi$ changes the nature of the instability qualitatively. $\delta>0$ as $q \to 0$.    }
    \label{fig2}
\end{figure}

\section{Details of the numerical simulations}

We numerically integrate the equation of motion using a pseudo-spectral method on a two-dimensional square periodic domain of side length $L$, discretized on an $N_x \times N_y$ uniform grid. In the simulations reported in the main paper, we start from random initial conditions and use $L=N_x=N_y=1024$ in the simulation shown in Fig. (1) and Fig. (3) and $L=N_x=N_y=512$ in the simulation shown in Fig. (2).

To reduce aliasing errors from the pseudo-spectral evaluation of the nonlinear term, we app standard 2/3-dealiasing rule in Fourier space. Modes satisfying

$$
\left|k_x\right| \geq \frac{2}{3} k_{\max } \quad \text { or } \quad\left|k_y\right| \geq \frac{2}{3} k_{\max }
$$

are filtered out in the nonlinear update.

We use a first-order exponential time-differencing scheme, ETD1, with a fixed time step $\Delta t=0.01$, for time integration. In this scheme, the linear part of the equation is integrated implicitly in Fourier space through the exponential propagator $e^{\Delta t \mathcal{L}}$, while the nonlinear term is treated explicitly using its value evaluated from the field at the current time step. Thus, the Fourier amplitudes are updated as

\begin{equation}
\phi_{\mathbf{k}}(t+\Delta t)=e^{\Delta t \mathcal{L}(\mathbf{k})} \phi_{\mathbf{k}}(t)+\Delta t \varphi_1(\Delta t \mathcal{L}(\mathbf{k})) \mathcal{N}_{\mathbf{k}}(t),
\end{equation}

Where $\varphi_1(z) =e^z-1/z $. The ETD formulation improves numerical stability by exactly resolving the linear growth while retaining an explicit pseudo-spectral evaluation of the nonlinear contribution.

\section{DESCRIPTION OF THE MOVIES}
\begin{itemize}
    \item \textbf{SMov1:} The movie shows intermittent target chaos. It captures the proliferation and evolution of target patterns, including the distortion of a target core before it collides with a motile boundary. After the collision, the target is replaced by traveling waves and several bound defect cores of opposite topological charge. The traveling-wave state is subsequently destroyed by renewed target proliferation; the cycle then repeats. We plot $\phi$ to emphasize the proliferation of the target core, and $\rho$ to visualize both the motile boundary and the distortion of the target core. As for the simulation parameters, the domain size $L=512$, the grid resolution $N=512$, $\alpha=0$, and $\beta=4$.
    \item \textbf{SMov2:} The movie shows cycles of creation and annihilation of topological defects and long-lived dislocation chaos. We see that both the defect clusters and topological defects are drifting. We plot $\phi$ to highlight the existence of defect clusters and lines of defects, and $\rho$ to show the movement of defect cores. As for the simulation parameters, the domain size $L=1024$, the grid resolution $N=1024$, $\alpha=0.6$, and $\beta=4$.
\end{itemize}

Supplemental movies are available \href{https://owncloud.gwdg.de/index.php/s/1axZz5Ca2sxGQKO}{here}.

\end{document}